\documentclass[pra,amsmath,amssymb,twocolumn,amsfonts]{revtex4}
 
\usepackage{amsbsy}
\usepackage{amsfonts}
\usepackage{amssymb}
\usepackage{amsmath}
\usepackage{graphicx,color}
\usepackage{graphicx}

\newcommand{\ad}{a^{\dagger}}

\newcommand{\ket}[1]{| #1 \rangle}
\newcommand{\bra}[1]{\langle #1 |}

\definecolor{gray}{rgb}{.4,.4,.4}
\definecolor{deepgreen}{rgb}{.1,.6,.3}

\begin{document}


\title{Observing different quantum trajectories in cavity QED}

\author{Marcelo Fran\c{c}a Santos}
\affiliation{1-Departamento de F\'isica, Universidade Federal de Minas Gerais, Belo Horizonte, Caixa Postal 702, 30123-970, MG, Brazil}
\affiliation{2-Centre for Quantum Technologies, National University of Singapore, 2 Science Drive 3, 117542 Singapore}

\author{Andre R. R. Carvalho}
\affiliation{Australian Centre for Quantum-Atom Optics, Department of Physics, Research School of Physics and Engineering, The Australian National University, ACT 0200, Australia}

\begin{abstract}
The experimental observation of quantum jumps is an example of single open quantum systems that, when  monitored, evolve in terms of stochastic trajectories conditioned on measurements results. Here we present a proposal that allows the experimental observation of a much larger class of quantum trajectories in cavity QED systems.  In particular, our scheme allows for the monitoring of engineered thermal baths that are crucial for recent proposals for probing entanglement decay and also for entanglement protection. The scheme relies on the interaction of a three-level atom and a cavity mode that interchangeably play the roles of system and probe. If the atom is detected the evolution of the cavity fields follows quantum trajectories and vice-versa.
\end{abstract}

\maketitle

\section{Introduction}
One of the most remarking properties of quantum mechanics is that the state of a quantum system changes not only via the deterministic evolution given by the Schr\"odinger equation but also when 
it is measured. Although the system can be measured directly, in many situations it is probed by an auxiliary system (ancilla) that is then detected, providing information about the system of interest. This is a typical situation in cavity QED where atoms and light interact and the detection of either of them alters the state of the other. In the microwave regime, for example, it is usually the atoms that behave as a probe for the field. This configuration has led to a number of experiments on fundamental aspects of quantum mechanics including the measurement of the decoherence of a cat state~\cite{Brune:1996}, QND measurement of single photons~\cite{Nogues:1999}, and the observation of quantum jumps of light~\cite{Gleyzes:2007}.

This property becomes particularly interesting for quantum open systems where the environment that the system is coupled to plays the role of a natural bona fide ancilla. The time evolution of a single quantum system can therefore be probed by directly monitoring this reservoir. This dynamics is going to be stochastic, as the evolution is conditioned on the measurements results, and can be mathematically described in terms of quantum trajectories, which are related to different physical ways to monitor the environment and to extract information about the system. 

The monitoring of the field leaving a damped cavity mode provides a good example of this stochastic dynamics. For instance, if a photodetector is used to collect the output of the cavity, the dynamics is better described in terms of quantum jumps where each click in the detector corresponds to lowering the number of photons inside the cavity by one. However, a completely different dynamics is found if the same propagating field is combined with a local oscillator in a beam splitter and a homodyne measurement is performed, in which case the time evolution of the damped cavity field is appropriately described by a continuous stochastic trajectory (quantum state diffusion). All these trajectories present very interesting scenarios that allow for the production of non-classical states of the cavity field as well as the protection of the purity of the cavity mode or of the entanglement shared by two or more modes. 
Note that if the lost photons are ignored, or the measurements averaged out, one obtains the usual master equation dynamics for the decoherence process that is responsible, for example, for rapidly turning superpositions of coherent states into mixtures~\cite{Brune:1996}. For this reason, each monitoring scheme described above is said to represent an unravelling of the master equation in terms of stochastic trajectories.

Note that the unravellings above represent only a limited set of the possibilities to measure the environment. In fact the master equation can be mathematically unravelled in infinite ways in terms of stochastic trajectories~\cite{Carmichael:1993,Rigo:1997,Wiseman:2000} and therefore one could envisage more general ways of monitoring the environment rather than only simple photodetection and homodyning. This freedom in defining unravellings has been recently explored in the context of entanglement decay and protection~\cite{Nha:2004,Carvalho:2007b,Viviescas:2010,Mascarenhas:2010a,Vogelsberger:2010, Mascarenhas:2010b, Mascarenhas:2010c, Yi:2010, Carvalho:2011}, where monitoring schemes that combine different decay channels play a crucial role on the recovery of the mixed state entanglement dynamics in terms of trajectories~\cite{Carvalho:2007b,Viviescas:2010, Vogelsberger:2010, Mascarenhas:2010b} and on the protection of entanglement conditioned on measurement outcomes~\cite{Carvalho:2011}. It would be interesting then to propose realistic experimental scenarios where this variety of unravellings could be explored. 

While the monitoring of the field emitted by a leaky cavity mode or by a decaying atom is within current experimental feasibility, this is not necessarily the most efficient way and certainly not the most complete one to generate different unravellings. 
This scheme proves to be very limited in two very interesting situations: non-zero reservoir temperatures and unravellings that combine detections and no-detections (jumps and no-jumps). The first case presents two problems: first how to determine when a very large reservoir loses a single photon to the system and second how to distinguish a photon that comes from the system from one that already exists in the reservoir. In the second case, one faces the problem of physically superposing a click with a no-click in the detector. These limitations severely hinder the exploration of quantum trajectories in these systems. 

Even though the direct monitoring of a natural thermal environment remains a challenge, in cavity QED one can artificially engineer this and other reservoirs that could produce different kinds of quantum trajectories when measured. In the microwave regime, for example, beams of atoms crossing the cavity can be used to mimic a thermal dissipative reservoir for the cavity field~\cite{Sargent-III:1974,Kist:1999,Pielawa:2010}. The posterior detection of these atoms produces quantum trajectories for the cavity field, as analyzed in~\cite{Kist:1999} both in terms of jumps and continuous diffusion processes. However, these schemes also present limitations to the production of different unravellings. For example, while the combination of different channels can be easily accomplished within the optical detection scenario by having the photons for each channel arriving at different ports of a beam splitter~\cite{Viviescas:2010,Carvalho:2011}, the situation for atomic detection seems far more complex. In all previous proposals to engineer thermal reservoirs for cavities using atomic beams, decay and excitation channels correspond to two-level atoms entering the cavity either in the ground or in the excited state. A combined detection, which is utterly important for the applications proposed in~\cite{Carvalho:2007b,Viviescas:2010,Mascarenhas:2010a,Vogelsberger:2010,Carvalho:2011}, would then require some kind of interaction between the atoms after they cross the cavity, which albeit not impossible, seems rather challenging.

In this paper we show that simple modifications to these previous proposals lead to an alternative scheme to engineer a thermal reservoir in the context of microwave cavity QED that accommodates more general detections, hence enabling the simulation of a wide set of different classes of trajectories for a monitored atomic reservoir. In particular, we address the limitations raised in the two previous paragraphs. We then propose a complementary experimental setup where we invert the roles, letting the cavity field play the reservoir for the atoms. We rely on the Purcell effect to channel the atomic decays into the modes of a lossy cavity and then use the optical detection proposed in~\cite{Carvalho:2011} to monitor the reservoir in different ways. In this last case, we introduce an additional measurement possibility to implement the proposal in~\cite{Carvalho:2011} that would circumvent the challenging task of collecting broadly emitted photons from atomic decay. 

\section{Atoms as the reservoir}

We will start by briefly recalling the usual way to simulate a thermal reservoir for a cavity field using a sequence of two-level atoms as discussed in~\cite{Sargent-III:1974,Kist:1999, Englert:2002}. The atoms, prepared in either the ground or excited states, cross a cavity where they interact resonantly via a Jaynes-Cummings Hamiltonian for a short enough time. By short time, we mean a time interval $\delta t$ much smaller than the inverse of the vacuum Rabi frequency of the system $1/\lambda$ such that the probability of an atom making a transition is small. 
Under this model, one can then show that if the atoms are not detected after interacting with the cavity, the cavity field (of mode annihilation operator $a$) will evolve according to a master equation of the form 
\begin{equation}
\label{eq:decoh_general}
\dot{\rho}=\gamma_{-} \mathcal{D}[a] \rho + \gamma_+ \mathcal{D}[a^\dagger] \rho,
\end{equation}
where $\mathcal{D}(c) \rho = c\rho c^\dagger -1/2 (c^\dagger c \rho +\rho c^\dagger c)$, and the relationship between the rates $\gamma_{-}$ and $\gamma_{+}$ is given by the ratio between the flux of atoms initially prepared in the ground ($r_g$) and excited state ($r_e$). A thermal bath is obtained by setting this ratio to be $r_e/r_g=\bar n/(1+\bar n)$, where $\bar n$ is the average number of photons in the modes with energy $\hbar \omega_c$ in the reservoir.

If one now considers that the atoms are detected after the interaction, then the evolution of the cavity field is conditioned on the result of this measurement and better described by quantum trajectories. 
In this case, if an atom enters in the ground (excited) state and is detected in the excited (ground) state then the cavity field will be modified by a jump operator $J_- = a$ ($J_+ = a^{\dagger}$) corresponding to the annihilation (creation) of a photon in the cavity. Note that these jumps will be rare since, under our assumption of small interaction time, the atoms will most probably remain in their original state, in which case the cavity field evolution will correspond to the application of a no-jump operator $J_0=(1-\frac{dt}{2}\sum_i J_i^{\dagger} J_i)$. If the state of the atom is known before and after the cavity then a quantum jump unravelling corresponding to the operators $\{J_0,J_-,J_+\}$ will be produced for the evolution of the cavity field~\cite{Kist:1999}. Note, as mentioned before, that in order to produce any other jump-like unravelling which necessarily combines the operators $J_-$ and $J_+$, the atoms would need to interact after going through the cavity which is a rather difficult experimental task. 

\begin{figure}[h]
\hspace{1cm}
\includegraphics[width=8.5cm]{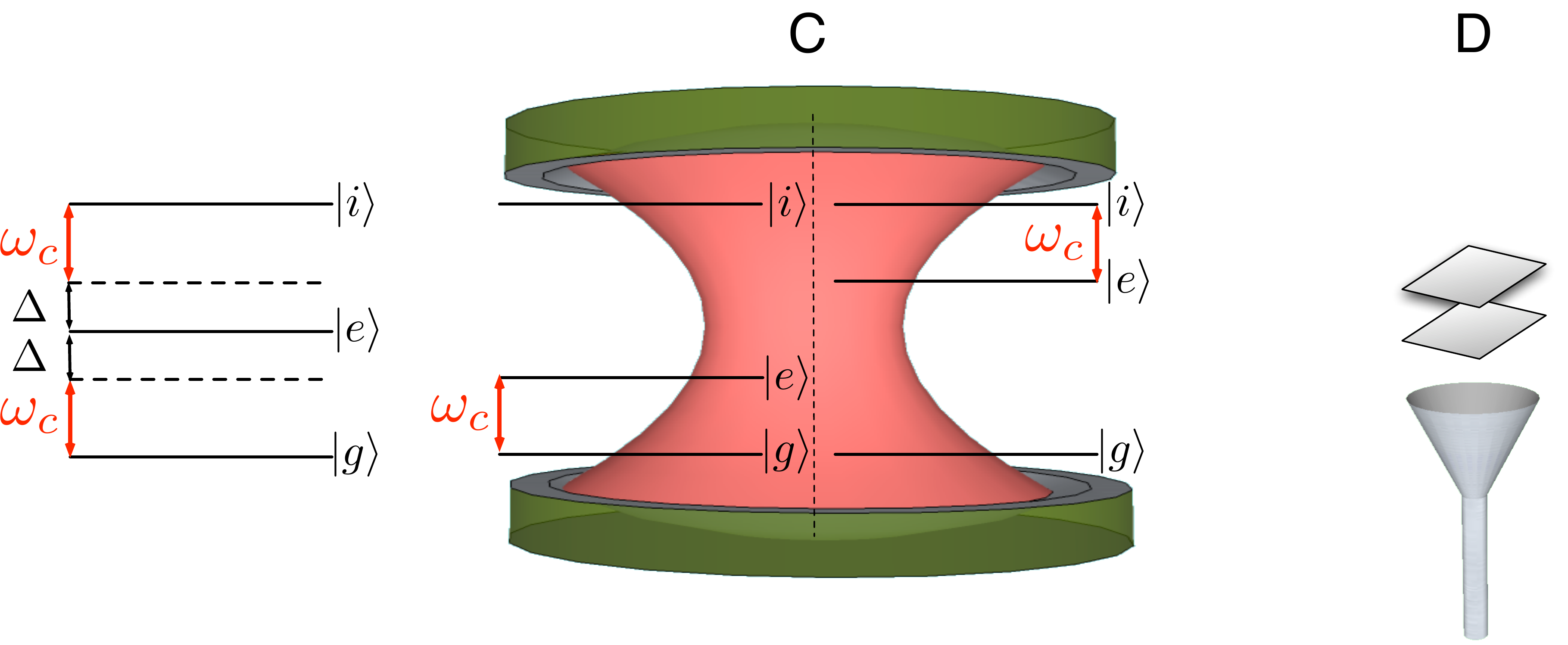}
\caption{Atoms with the level scheme shown on the left cross the cavity C before being detected. Inside the cavity, the application of external fields change the atomic levels in two stages: first the cavity mode becomes resonant with the $e-g$ transition, then with $e-i$ transition. Rotations between levels $i$ and $g$ after the interaction with the cavity and before atomic detection produces different unravellings of the dynamics in terms of quantum jumps.}
\label{fig1}
\end{figure}

We now present an alternative scheme to engineer a thermal reservoir that naturally allows more general detections. The central idea is to encapsulate in a single atom the interactions with the cavity that correspond to the evolution with $J_-$ and $J_+$. For that we will use three-level atoms with levels organised in a cascade, as shown in Fig. \ref{fig1}, and selected in such a way that the energy difference between the intermediate level ($|e\rangle$) and the lower and upper levels (respectively $|g\rangle$ and $|i\rangle$) are close enough to the energy of the chosen cavity mode so that these transitions can be tuned into resonance through the application of external fields, but only one at a time. The atoms are prepared in the $\ket{e}$ state before entering the cavity and in the first stage, the external field shifts the $|e\rangle$ level such that the transition  $|g\rangle \rightarrow |e\rangle$ becomes resonant with the cavity frequency $\omega_c$. The system then evolves under the Jaynes-Cummings Hamiltonian $H_1= \hbar \lambda_1 (\ket{e}\bra{g} a +\ket{g}\bra{e}\ad)$ for a short time $\delta t_1$. Up to second order on $\lambda_1 \delta t_1$ the initial atom-field state $\ket{\Psi(t)}=\ket{e}\ket{\Phi}$ will be transformed to
\begin{eqnarray}
&\ket{\tilde \Psi(t+\delta t_1)}=\left(\ket{e}-\frac{(\lambda_1 \delta t_1)^2}{2} a a^\dagger \ket{e} -i \lambda_1 \delta t_1 \, a^\dagger \ket{g} \right) \ket{\Phi}, \nonumber \\
&
\label{eq:psi1}
\end{eqnarray}
where $\lambda_1$ is the coupling constant, $|\Phi\rangle$ the initial field state and the tilde indicates that the state is not normalised. If, now, the other transition $|e\rangle \rightarrow |i\rangle$ is tuned to resonance by changing the external field, the Hamiltonian will be $H_2= \hbar \lambda_2 (\ket{i}\bra{e} a +\ket{e}\bra{i}\ad)$ and, again, making the short time expansion will lead to 
\begin{eqnarray}
\ket{\tilde \Psi(t+\delta t_1+\delta t_2)}=\left(\ket{e}-i \lambda_1 \delta t_1 \, a^\dagger \ket{g} -i \lambda_2 \delta t_2 \, a \ket{i}  \right. \nonumber \\
\left. -\frac{(\lambda_1 \delta t_1)^2}{2} a a^\dagger \ket{e}  -\frac{(\lambda_2 \delta t_2)^2}{2} \ad a \ket{e} \right) \ket{\Phi}.
\label{eq:psi2}
\end{eqnarray}

Considering now a time interval $\Delta t$ that is large enough so that $n$ atoms ($n=r \Delta t \gg 1$) cross the cavity and yet the probability of occurrence of a quantum jump remains very small, one can recover the master equation result by tracing out the atomic system. This becomes more evident if one identifies the rates as $\gamma_+ = r (\lambda_1 \delta t_1)^2$ and $\gamma_- = r (\lambda_2 \delta t_2)^2$, where $r$ is the number of atoms entering the cavity per unit of time. Therefore, the same atom can emulate at the same time both the dissipative and the excitation reservoirs. 

If instead of tracing out the atoms one measures their levels, then the evolution of the state of the cavity field obeys a quantum trajectory given by the no-jump operator $J_0=(1-\frac{\gamma_+ \Delta t}{2} a a^\dagger-\frac{\gamma_- \Delta t}{2} a^\dagger a)$ if the atom is detected in level $|e\rangle$ or the jump $J_- = \sqrt{\gamma_- \Delta t} \, a$ ($J_+ = \sqrt{\gamma_+ \Delta t} \, a^\dagger$) if the atom is detected in level $|i\rangle$ ($|g\rangle$), that corresponds to the usual monitoring discussed previously.

The advantage of encoding both jumps in the same atom is that it allows us to explore different unravellings by introducing simple modifications in the detection scheme. If one makes a unitary transformation between levels $\ket{i}$ and $\ket{g}$ after the atoms cross the cavity, for example, then a detection of each state will correspond to a different unravelling in which the jump operators are combinations of $a$ and $a^\dagger$. For instance, in the particular case that $\gamma_- = \gamma_+$, a $\pi /2$ pulse between levels $|g\rangle$ and $|i\rangle$ produces jump operators proportional to the field quadratures $a+e^{i\phi}a^\dagger$ and $a-e^{-i\phi}a^\dagger$ where the phase $\phi$ is given by the chosen rotation. In fact, since one can apply unitary operations also involving level $|e\rangle$ then it is clear that any unravelling of the type $UJ$, where $U$ is a unitary matrix acting on the vectorial space of the jumps and $J$ is the vector $J=(J_0,J_+,J_-)$, can be produced. Note that the unitary $U$ is obtained by individually rotating the atoms after they have already interacted with the cavity mode therefore preserving the master equation structure when the atoms are ignored. 

In the scheme described above, the detection of the atom modifies the quadratures of the field affecting, in principle, all Fock states. For example, if the field is initially constrained to Fock states $\{|0\rangle,|1\rangle\}$, the detection of the atom in its middle ($|e\rangle$) or highest ($|i\rangle$) levels will preserve the initial subspace, but the detection of the atom in its ground state will expand the effective subspace of the cavity mode to include Fock state $|2\rangle$. This can be avoided by choosing a selective interaction~\cite{Marcelo:2001} for the $|e\rangle \rightarrow |g\rangle$ transition in which case the creation of an extra photon in the cavity can be set to occur only if the cavity is empty, hence preserving the initial effective subspace and thus fully mimicking a spin-half system for the cavity field~\cite{Marcelo:2005}.

\section{Cavity fields as reservoirs}

\begin{figure}[ht]
\includegraphics[width=8.5cm]{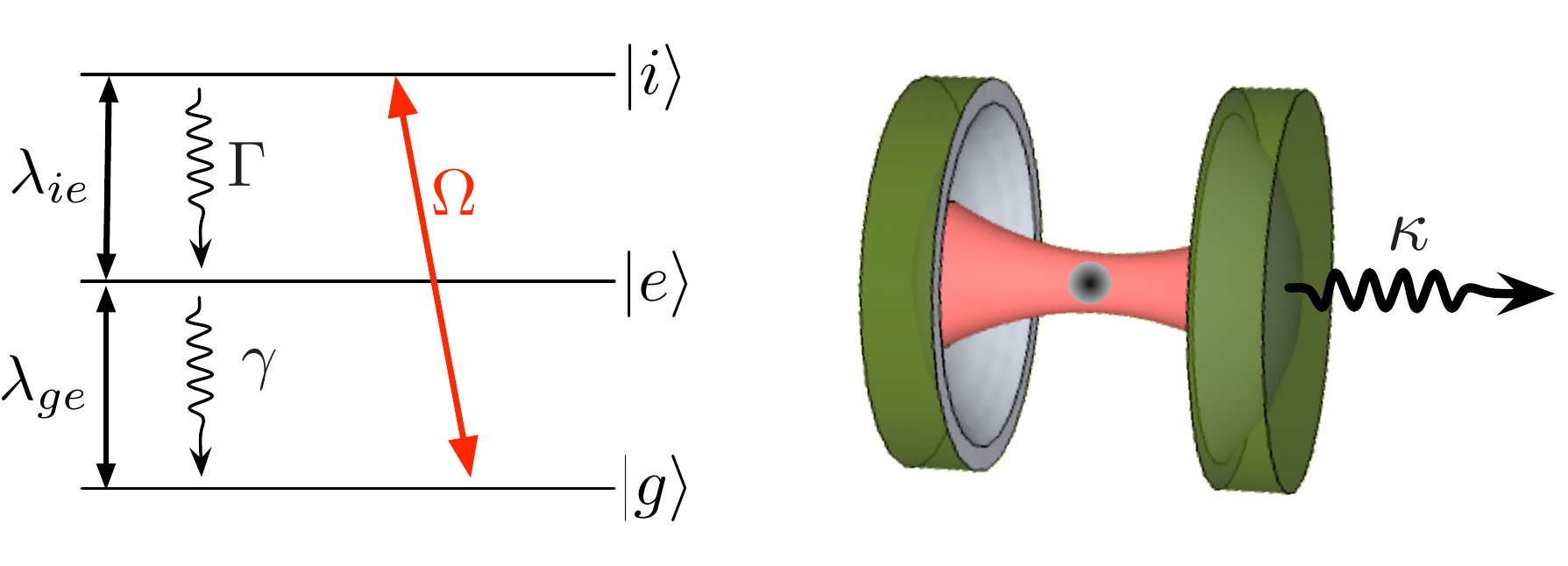}
\caption{An atom with the level scheme shown on the left is placed inside a cavity and is coupled to two orthogonal modes of a damped cavity (right panel). The atom-cavity couplings ($\lambda_{ge}$ and $\lambda_{ie}$) and the cavity decay ($\kappa$) will induce an effective atomic decay via the Purcell effect. Photons from the right and left polarized modes leaving the cavity will then correspond to $\ket{e}\rightarrow \ket{g}$ and $\ket{i}\rightarrow \ket{e}$ transitions, respectively. The corresponding effective decay rates $\gamma_{eg}$ and $\gamma_{ie}$ are assumed to be much larger than the natural spontaneous decay rates $\gamma$ and $\Gamma$.}
\label{fig2}
\end{figure}

Inverting the roles of the previous session, now, the three level atom is stationary and interacts with two orthogonal cavity modes such that the $|i\rangle \rightarrow |e\rangle$ ($|e\rangle \rightarrow |g\rangle$) transition generates photons circularly polarized to the Right (Left) associated to the annihilation operator $a_R$ ($a_L$) of the cavity mode. We assume a very lossy cavity (of decay rate $\kappa$) made of a nearly perfect mirror and a semi-transparent one (see Fig.~\ref{fig2}). If the Rabi frequencies of each transition are large enough then the Purcell effect will channel the atomic decay in the cavity modes, and the cavity asymmetry will guarantee that the photons will always leak in a well defined direction. 

Each atomic transition has to be coupled to its respective cavity mode with different coupling constants $\lambda_{ie}$ and $\lambda_{eg}$, and a $\pi$-polarized classical field of strength $\Omega$ has to drive the $|g\rangle \rightarrow |i\rangle$ transition as shown in Fig.~\ref{fig2}. The relation between the involved rates should be $\kappa \gg \lambda_{ie} > \lambda_{eg} > \Omega \gg \gamma,\Gamma$, where $\gamma$ and $\Gamma$ are the natural atomic linewidths. For the scheme to work, much greater typically reads $(\kappa, \lambda_{ie}, \lambda_{eg}, \Omega, \gamma, \Gamma)=(1000, 100, 5, 1, 0.01, 0.01)$ where this particular choice has already been made to fit the analysis that follows.

The overall master equation that describes the dynamics of the whole system is given by:
\begin{eqnarray}
\label{me_total}
\dot{\rho}= -\frac{i}{\hbar}[H,\rho] + \gamma \mathcal{D}[\sigma_-^{(eg)}]\rho + \Gamma \mathcal{D}[\sigma_-^{(ie)}]\rho + \kappa \mathcal{D}[a],
\end{eqnarray}
where 
\begin{eqnarray}
\frac{H}{\hbar}=\omega_{e}|e\rangle \langle e|+ \omega_i |i\rangle \langle i| + \omega_R a_R^\dagger a_R + \omega_L a^\dagger_L a_L \nonumber
\\+ i\lambda_{ge} (a_L \sigma_+^{(eg)}-a_L^\dagger \sigma_-^{(eg)}) + i\lambda_{ei}  (a_R \sigma_+^{(ie)}-a_R^\dagger \sigma_-^{(ie)}) \nonumber 
\\+ i(\sigma_+^{(ig)} e^{i\omega_{i} t}-\sigma_-^{(ig)} e^{-i\omega_{i} t}),
\end{eqnarray}
with $\sigma_-^{ij}$ and $\sigma_+^{ij}$ being the lowering and raising operators for the $ij$ transition.
Being $\kappa \gg \lambda_{ge}$ there is no Rabi oscillation between the corresponding atomic transition and the cavity mode which acts just as a channel for the atomic decay. In other words, as soon as a photon is transferred from the atom to the cavity mode it leaks through the semi-transparent mirror. The effective atomic decay rate is then given by $\gamma_-=\gamma_{eg}=\frac{4\lambda_{ge}^2}{\kappa}$ as long as $\gamma \ll  \frac{4\lambda_{ge}^2}{\kappa}$. Much in the same way, the effective decay rate for the $|i\rangle \rightarrow |e\rangle$ transition is given by $\gamma_{ie}=\frac{4\lambda_{ie}^2}{\kappa}$. Now, if the external driving field $\Omega$ is still much smaller than this effective decay rate, then level $|i\rangle$ can be adiabatically eliminated and the combined effect of the classical field and the strong decay rate $\gamma_{ie}$ will be to generate an effective incoherent pump between levels $|g\rangle$ and $|e\rangle$ of rate $\gamma_+=\frac{4\Omega^2}{\gamma_{ie}}=\frac{\Omega^2 \kappa}{\lambda_{ie}^2}$. In this way, the scheme produces a decay and an excitation reservoir for the qubit composed of levels $e$ and $g$ with rates $\gamma_-$ and $\gamma_+$, respectively. In particular, when the effective rates $\gamma_{\pm}$ are the same, \textit{i.e.} $\frac{\Omega^2 \kappa}{\lambda_{ie}^2} = \frac{4\lambda_{ie}^2}{\kappa}$, one obtains an infinite temperature environment.

Note that there is still another degree of freedom available if the conditions are not perfectly matched that is the eventual detuning between cavity modes and atomic transitions. For example, let us say that $\lambda_{ge} \sim \lambda_{ei}$. If $\Delta_{ge} = \omega_e - \omega_L > 0$, then the effective decay rate is corrected by $\gamma_{-} =  \frac{4\lambda_{ge}^2}{\kappa (1+4\Delta_{ge}^2 /\kappa^2)}$~\cite{AlexiaMarcelo}.

Under the above approximations and assuming that the effective rates are much larger than the natural atomic linewidths ($\gamma_{\pm} \gg \gamma, \Gamma$), the unconditioned dynamics of the (two-level) atom can be finally written as:
\begin{eqnarray}
\dot{\rho}\approx \gamma_- \mathcal{D}[\sigma_-]\rho + \gamma_+ \mathcal{D}[\sigma_+]\rho,
\end{eqnarray}
where the superscript $(eg)$ has been dropped for simplicity.
If the photons leaving the cavity are detected, then the atomic dynamics will follow a stochastic dynamics conditioned on the measurement outcomes as described by the quantum trajectory approach. Equivalently to the previous session, different detection techniques will lead to different unravellings. For example, if a $\lambda /4$ plate is placed right after the leaking mirror, then each circular polarization is converted into orthogonal linear polarizations, $a_R \rightarrow a_H$ ($a_L \rightarrow a_V$). These photons are then sent into a polarized beam splitter that separates the linear polarizations into two different propagation modes where detectors are placed to collect them. In this way, the click in the $a_H$ ($a_V$) path identifies a transition $|g\rangle \rightarrow |e\rangle$ ($|e\rangle \rightarrow |g\rangle$) which corresponds to the ``usual'' photodetection unravelling with the jumps $J_+=\sqrt{\gamma_+ dt} \, \sigma_+^{(eg)}$ and $J_-= \sqrt{\gamma_- dt}\, \sigma_-^{(eg)}$ applied to the atomic state. If no photon is detected in the $dt$ interval, then the no-jump $J_0=(1-\frac{\gamma_-dt}{2} \sigma_{ee} -\frac{\gamma_+ dt}{2} \sigma_{gg})$ operator is applied to the system.

A more interesting situation occurs if the $\lambda /4$ plate is taken out of the setup. In this case, the photons propagating after the PBS will correspond to linear combinations of the photons leaking from the atom, implementing an unravelling with the jumps $\sigma_x=\sigma_-^{(eg)} + \sigma_+^{(eg)}$ and $\sigma_y=\sigma_-^{(eg)} - \sigma_+^{(eg)}$ (similar to the quadrature unravelling described in the previous section). We have shown in~\cite{Carvalho:2011} that if two qubits initially share an entangled state, then this kind of monitoring performed on both subsystems can preserve the entanglement in the system. The above setup is therefore a way to implement the entanglement protection ideas proposed in~\cite{Carvalho:2011}. 

Before concluding, let us just remark that the scheme here described for cavity QED could also be applied to other experiments involving three level systems and harmonic oscillators. Natural candidates where quantum jumps have been recently observed or proposed are superconducting qubits~\cite{Han:2008, Siddiqi:2010} and nanoresonators~\cite{Milburn:2010a, Milburn:2010b}.

Finally, in conclusion, we have proposed two experimental ways to produce a wide range of unravellings of a master equation evolution in cavity QED systems. In the first case, three-level atoms are used to simulate a thermal dissipative reservoir and the preparation and posterior detection of these atoms in different basis produce unravellings that correspond to different combinations of the natural jump operators and even between these and the no-jump one. This is a particular case in which ``no-click'' and ``click'' can be physically combined to generate a new class of unravellings. Later, we also show how to invert the roles and use the Purcell effect to channel the detection of the spontaneous decay and of the incoherent pump (or even their combination) for driven three-level atoms. This work adds to the existing class of proposed cavity QED experiments, expanding the range of quantum effects that can be explored in this experimentally successful system. 

\acknowledgements
We would like to acknowledge the support from Center for Quantum Technologies at the National University of Singapore. ARRC acknowledges financial support by the Australian Research Council Centre of Excellence program and MFS acknowledges the support of the National Research Foundation and the Ministry of Education of Singapore.

\end{document}